\documentstyle[prb,aps,floats,graphicx,epsf,psfrag,placeins,subfigure]{revtex}
\twocolumn

\draft

\newcommand{\mub}{{$\mu_{B}$ }}
\newcommand{\mubp}{{$\mu_{B}$}}

\newcommand{\Ca}{CaMnO$_3$ }
\newcommand{\Cap}{CaMnO$_3$}

\newcommand{\La}{LaMnO$_3$ }
\newcommand{\Lap}{LaMnO$_3$}

\newcommand{\ABfive}{A$_{0.5}$B$_{0.5}$MnO$_3$ }

\newcommand{\AxByp}{A$_{1-x}$B$_{x}$MnO$_3$}

\newcommand{\LaCax}{La$_{1-x}$Ca$_x$MnO$_3$ }

\newcommand{\LaCa}{La$_{0.5}$Ca$_{0.5}$MnO$_3$ }

\newcommand{\PrCasix}{Pr$_{0.60}$Ca$_{0.40}$MnO$_3$ }

\newcommand{\LaCathree}{La$_{0.33}$Ca$_{0.67}$MnO$_3$ }
\newcommand{\LaCathreep}{La$_{0.33}$Ca$_{0.67}$MnO$_3$}

\newcommand{\LaCap}{La$_{0.5}$Ca$_{0.5}$MnO$_3$}

\newcommand{\NaVO}{NaV$_2$O$_5$ }
\newcommand{\Mnoct}{MnO$_{6}$ }

\newcommand{\eg}{e$_g$ }
\newcommand{\tg}{t$_{2g}$ }

\newcommand{\ooneion}{O$^{-}$ }
\newcommand{\ooneionp}{O$^{-}$}
\newcommand{\oion}{O$^{2-}$ }

\newcommand{\mmm}{Mn$^{3+}$ }
\newcommand{\mmmp}{Mn$^{3+}$}
\newcommand{\mmmm}{Mn$^{4+}$ }

\newcommand{\dt}{d$^{3}$ }
\newcommand{\df}{d$^{4}$ }
\newcommand{\dfp}{d$^{4}$}

\begin{document}

\bibliographystyle{prsty}

\title{Ferromagnetic polarons in \LaCa and \LaCathree}
\author{G. Zheng and C.H. Patterson}
\address{Department of Physics and Centre for Scientific Computation,\\
University of Dublin, Trinity College, Dublin 2, Ireland.}
\date{\today}
\maketitle

\begin{abstract}

Unrestricted Hartree-Fock calculations on \LaCax (x = 0.5 and x = 0.67) in 
the full magnetic unit cell show that the magnetic ground states of these compounds 
consist of 'ferromagnetic molecules' or polarons ordered in herring-bone patterns.  Each polaron 
consists of either two or three Mn ions separated by \ooneion ions with a magnetic moment 
opposed to those of the Mn ions.  Ferromagnetic coupling within the polarons is 
strong while coupling between them is relatively weak.  Magnetic moments on 
the Mn ions range between 3.8 and 3.9 \mub in the x = 0.5 compound and moments on the \ooneion
ions are -0.7 \mubp.  
Each polaron has a net magnetic moment of 7.0 \mubp, in good agreement with recently 
reported magnetisation measurements from electron microscopy.  The polaronic 
nature of the electronic structure reported here is obviously related to the Zener 
polaron model recently proposed for \PrCasix on the basis of neutron 
scattering data.

\end{abstract}


\section{Introduction}

\begin{figure}[h!]
\begin{center}
\includegraphics[height=14 cm]{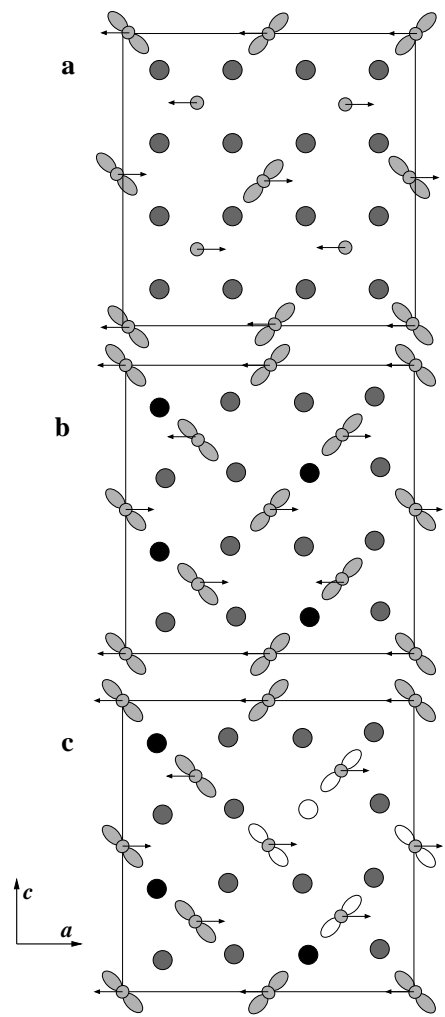}
\caption{Schematic illustrations of the magnetic unit cell for \ABfive
compounds with a CE magnetic structure: (a) the conventional double exchange 
picture; (b) the polaronic picture; (c) a possible change in orbital order in a small 
ferromagnetic domain when an electron is added at the oxygen ion site indicated by an 
unshaded circle.  Spins are indicated by arrows, Mn \df sites by double-lobed orbitals, 
Mn \dt sites by small shaded circles, \oion ions by larger shaded circles and \ooneion ions by 
black circles.}
\label{fig:fig1}
\end{center}
\end{figure}

The current paradigm for the electronic structure of manganites (\AxByp) with 
x $\geq$ 1/2 is a lattice of Mn sites with \dt or \df orbital occupancy, with the proportion 
of each decided by the value of x and orbital ordering (OO) of the occupied \eg orbital 
on \df sites.  The corresponding double-exchange model Hamiltonian has been 
studied extensively \cite{Yunoki98,Fratini00,Dagotto00}.  
Manganites with x $\geq$ 1/2 usually exhibit a phase transition \cite{Kawano97,Radaelli97,Radaelli99,Kim00,Kajimoto01,Pissas02}
which has been assumed to be charge ordering (CO) of the \dt and \df sites at a 
temperature well below the paramagnetic transition temperature.  The conventional picture of the orbital and 
charge ordered phase with the CE magnetic structure \cite{Wollan55} is shown schematically in Fig. \ref{fig:fig1}a.

The validity of this picture for the manganites has been questioned recently \cite{Millis95,Zhao00}; its contradictions  
were pointed out by Goodenough in 1955 \cite{Goodenough55}. An alternative picture, the Zener polaron \cite{Khomskii97}, which challenges the 
ideas of double-exchange and CO in the manganites, was recently proposed to account for
the structure of \PrCasix determined by neutron scattering \cite{Daoud02}.  
In the conventional CO and OO picture, ordering of the \eg electron on \df Mn sites is 
expected to induce a Jahn-Teller (JT) distortion; the long Mn-O bond distance in 
\La exceeds 2.15 \AA  \cite{Wollan55,Moussa98}  while the Mn-O distance in cubic perovskite \Ca is 
1.88 \AA \cite{Wollan55}.  The lesser modulations of bond length in doped manganites with x $\sim$ 1/2 
(1.92 and 2.06 \AA { in} \LaCa \cite{Radaelli97} and 1.98 and 2.01 \AA { in} \PrCasix \cite{Daoud02})
suggest an intermediate valence.   

\begin{figure}[h!]
\begin{center}
\includegraphics[height=4 cm]{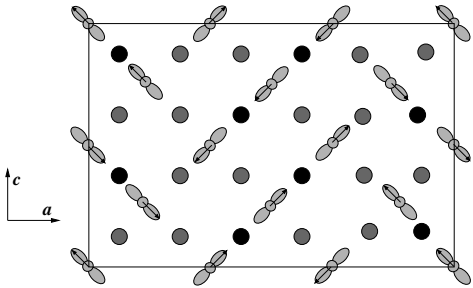}
\caption{Schematic illustration of the magnetic unit cell for \LaCathree
with the experimentally observed magnetic structure. Spins are indicated 
by arrows, Mn \df sites by double-lobed orbitals, \oion ions by shaded circles and \ooneion 
ions by black circles.}
\label{fig:fig2}
\end{center}
\end{figure}

In this Communication we report results of Unrestricted Hartree-Fock (UHF) 
calculations on \LaCax for x = 1/2 and x = 2/3.  The UHF electronic 
structure for these compounds is interpreted in terms of ferromagnetic 
polarons containing two (x = 1/2) or three (x = 2/3) Mn ions.  The main difference 
between results from UHF calculations and the conventional double exchange picture 
is that all Mn ions are essentially \dfp. Consequently  electrons must transfer from oxygen ions
to every other Mn ion in the x = 1/2 compound, resulting in an ordered array of \ooneion ions between
pairs of Mn ions.  The structures of the polaron phases for x = 1/2 and x = 2/3 are shown 
schematically in Figs. \ref{fig:fig1}b and \ref{fig:fig2}.  

Mn-\ooneionp-Mn  
and Mn-\ooneionp Mn-\ooneionp Mn chains constitute 
the polarons for x = 1/2 and x = 2/3, respectively.  Mn magnetic moments are in the range 3.8 to 3.9 \mub  
for either compound while
oxygen ions within the polarons have charges closer to \ooneion than \oion and magnetic moments of -0.7 \mubp.  
All other oxygen ions are 
essentially \oion ions.  Each polaron has a net magnetic moment of 7.0 \mubp, in good 
agreement with recently reported magnetisation measurements from electron 
holography and Fresnel imaging\cite{Loudon02}.  

UHF calculations predict that magnetic coupling within polarons is strong and 
ferromagnetic (FM) while coupling between polarons is much weaker and can be 
antiferromagnetic (AF) or FM; these observations lead to a natural explanation for the 
observed CE or A-type AF magnetic ground states of manganites with x = 1/2 \cite{Kawano97,Kajimoto01} and 
the magnetic ground state of \LaCathreep, which contains polarons with the 
magnetic moment roughly aligned along the polaron axis \cite{Radaelli99}.  Given that the magnetic coupling 
within the polarons is strong and FM, the appropriate model Hamiltonian for these 
systems at low temperature is a Heisenberg Hamiltonian on a triangular lattice where each polaron is a 
single magnetic entity, rather than the conventional double exchange Hamiltonian.  

Results of UHF our calculations are at odds with density functional 
theory (DFT) calculations in the local spin density approximation (LSDA) reported 
recently \cite{Popovic02,Medvedeva01} and with LSDA calculations done as part of this work using the same 
electronic structure code \cite{Crystal03} 
in that UHF calculations predict a magnetic moment on oxygen ions within 
the polarons whereas DFT calculations do not.  

\section{UHF Calculations}

All electron UHF calculations were performed for \LaCa and 
\LaCathree using the low temperature structures refined using x-ray 
data by Radaelli and coworkers \cite{structures} in the full (80 and 120 atom) magnetic 
unit cells.  

\begin{figure}[h!]
\begin{center}
\includegraphics[height=6 cm]{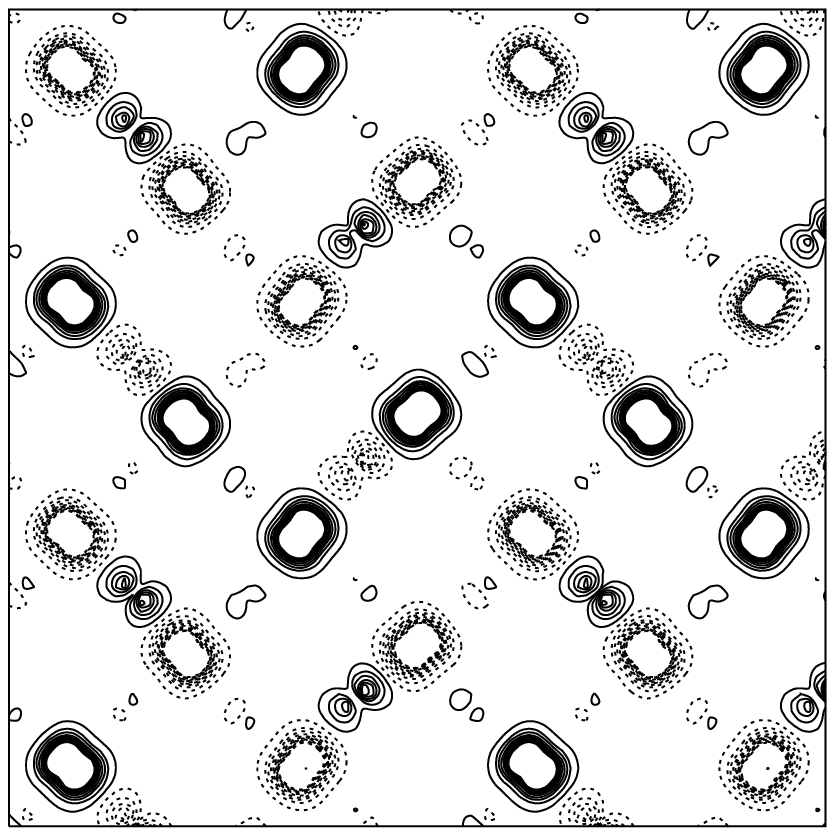}
\caption{UHF spin density plot for \LaCa in the CE-AF state.} 
\label{fig:fig3}
\end{center}
\end{figure}

\begin{figure}[h!]
\begin{center}
\includegraphics[height=6 cm]{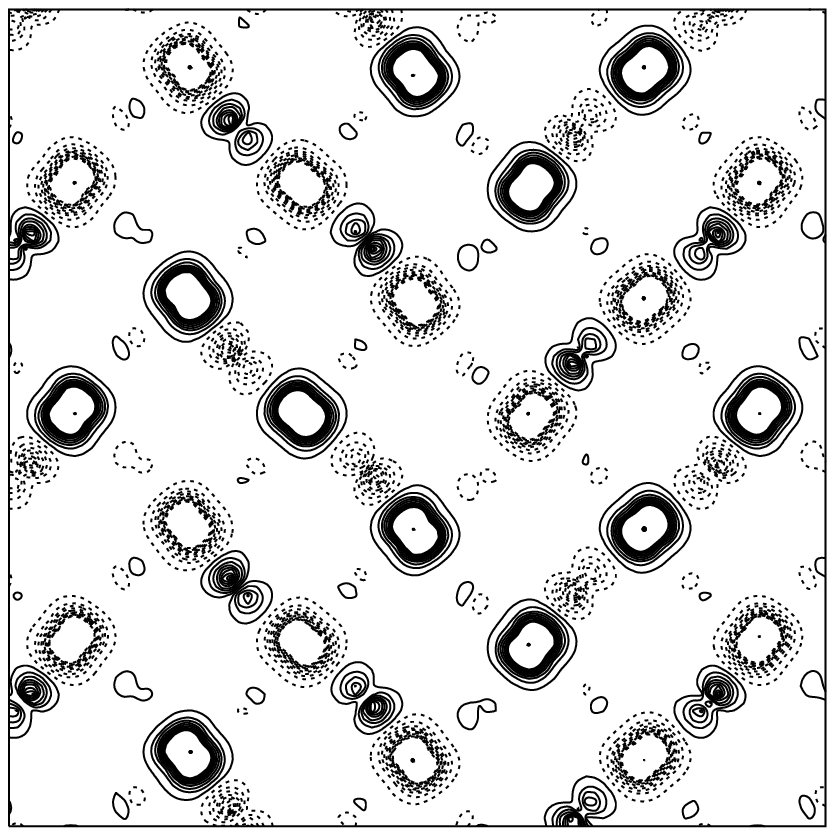}
\caption{UHF spin density plot for \LaCathree in AF state.} 
\label{fig:fig4}
\end{center}
\end{figure}

Spin densities of \LaCa (Fig. \ref{fig:fig3}) and \LaCathree (Fig. \ref{fig:fig4}) in the ac planes 
of the crystal structures clearly show the polaronic nature of the electronic structures, including
the magnetic moment of the \ooneion ions opposed to those of the neighbouring Mn ions.
The magnetic unit cells of either compound each contain four polarons in the ac planes shown.

Magnetic moments and 
Mulliken populations on Mn and oxygen ions in the polarons are given in Table \ref{tab:tab1}.  

\begin{table}[h!]
\caption{Charges and magnetic moments in \LaCa from Mulliken populations and experiment.}
\begin{tabular}{c d d}
                        
                &       Mn moment \mub  &               O moment \mub\\
\hline
UHF (CE-AF)\tablenotemark[1]            &		3.78, 3.91	&	-0.67\\
Expt.\tablenotemark[1]\tablenotemark[2] &       	2.57, 2.98	&	    -\\
Expt.\tablenotemark[3] 	                &               3.4 $\pm$ 0.2   &           -\\ 
\hline
                &       Mn charge       &               O charge\\
\hline
UHF (CE-AF)\tablenotemark[1]		&               2.18, 2.17	&	-1.24\\
\end{tabular}\label{tab:tab1}
\tablenotemark[1]{Two values for the Mn moment are quoted as the Mn ions in the polaron are inequivalent}\\
\tablenotemark[2]{Neutron scattering, Ref. \cite{Radaelli97}}\\
\tablenotemark[3]{Electron holography, value obtained for FM domain per Mn ion, Ref. \cite{Loudon02}}\\
\end{table}

Electron holography and Fresnel imaging\cite{Loudon02} measurements show that below the AF N\'eel temperature the 
x = 1/2 compound actually consists of both FM and AF domains
and that the magnetic moment per Mn ion in the FM domains was 3.4 $\pm$ 0.2 \mubp. 
UHF calculations predict a net magnetic moment of 7.0 \mub for the polaron, i.e. the polarons are fully spin polarised,
in agreement with these measurements of the magnetisation of FM domains \cite{Loudon02}.
Earlier neutron scattering data for this compound (Table \ref{tab:tab1}) indicated magnetic moments 
somewhat smaller than these fully polarised values \cite{Radaelli97} but the trend, where larger moments are found on the sites denoted
\mmmp,  is reproduced.

Mulliken populations from UHF calculations on \LaCax indicate
total Mn ion populations in the range 2.25 (x = 0) to 2.13 (x = 1) with a monotonic 
variation for intermediate values of x.  Mn d populations have an even smaller relative 
variation across the range of x with a d population of 4.66 in \Lap, a range of d populations from 4.72 to 4.73 
in \LaCap, and 4.70 in \Cap.
This is consistent with a bonding picture in which the \tg shell on each Mn ion is half-filled 
and a pair of \eg orbitals is combined with a set of four empty sp$^{3}$ orbitals to form a set of 
equivalent d$^{2}$sp$^{3}$ octahedral hybrid orbitals for polar-covalent Mn-O bonding.  The 
consistency of both the Mn ion population and d population across the range of values of x contradicts the 
conventional double exchange picture where Mn ions are assumed to have their 
formal \mmm and \mmmm charges.

Total energies were calculated for the FM state and four different low energy AF 
states of the x = 1/2 compound. Low energy states are found by flipping entire polaron magnetic moments; flipping the spin of
one of the Mn ions in a polaron results in an increase of the total energy by around 400 meV.
Energies of the various low energy magnetic states of the x = 1/2 compound are 
well-fitted by an Ising-like, nearest-neighbour Hamiltonian of the form given in Eq. \ref{eqn:eqn1}.

\begin{equation}\label{eqn:eqn1}
H = \sum_{<ij>}{}  J_{ij} \frac{\hat{S}_{zi}.\hat{S}_{zj}}{S^{2}}
\end{equation}

\begin{figure}[h!]
\begin{center}
\includegraphics[height = 3 cm]{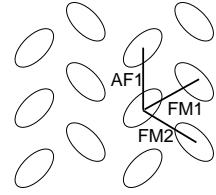}
\caption{Schematic illustration of exchange couplings between polarons in a 
Heisenberg Hamiltonian description.}
\label{fig:fig5}
\end{center}
\end{figure}

Labelling of exchange couplings is
shown schematically in Fig. \ref{fig:fig5} and exchange constants obtained by fitting total 
energies of UHF calculations are given in Table \ref{tab:tab2}. The spin magnitude in Eq. \ref{eqn:eqn1} was chosen to be \textit{S} = 2. 

Exchange couplings within the plane shown in Fig. \ref{fig:fig5} along zig-zag chains are FM and roughly equal 
in magnitude whereas they are AF perpendicular to the zig-zag chains.
Exchange coupling between polarons in different planes is AF. The magnetic ground state
is expected to be CE-type when J$_{AF1}$ exceeds (J$_{FM1}$ + J$_{FM2}$)/2 in magnitude and A-type otherwise.
Both are observed to be the magnetic 
ground state for various x = 1/2 compounds, depending on the A and B ion types \cite{Kajimoto01}.  

\begin{table}[h!]
\caption{Exchange coupling constants derived from \LaCa UHF calculations.}
\begin{tabular}{c d d}
               	               
                &      Exchange constant (meV)\\
\hline
J$_{AF1}$			&	  5\\	
J$_{AF2}$\tablenotemark[1]	&	  8\\	
J$_{FM1}$			&	-14\\	
J$_{FM2}$			&	-12\\	
\end{tabular}\label{tab:tab2}
\tablenotemark[1]{Average value over two interplanar couplings in crystallographic unit cell}
\end{table}

Our calculations actually predict an A-type AF ground state whereas the CE-AF structure is the 
ground state in \LaCa \cite{Wollan55}.   However, in similar UHF calculations on 
\La \cite{Nicastro02}, it was found that AF couplings are significantly underestimated compared to 
either experiment or results of configuration interaction (CI) cluster calculations, 
whereas FM couplings are in better agreement with both.  

\section{Discussion}

The polaronic picture for manganites with x $\geq$ 1/2 which has been presented is 
consistent with a wide range of experimental data and allows various observations, 
such as the unusual magnetisation in \LaCathreep, to be explained:  it 
accounts for the observation of A-AF or CE-AF ground states for various 
combinations of counterion in \ABfive \cite{Kawano97,Kajimoto01}; it is consistent with full spin 
polarisation of Mn ions in \LaCa \cite{Loudon02}.

One must be cautious in using UHF calculations to estimate magnitudes of magnetic 
moments on the oxygen ions.  An analogy can be drawn between electron transfer from 
the O ion in a polaron to a neighbouring Mn ion and separation of the electron pair 
in a hydrogen molecule as the proton-proton distance is increased above the molecular 
equilibrium bond length: UHF wave functions for hydrogen molecules at large bond 
distances consist of a spin-up electron on one proton and an spin-down electron on the 
other; the additional configuration in which spins on protons have been exchanged is 
omitted, owing to the simple functional form of the wave function.  UHF (and LSDA) 
wave functions incorporate strong electron correlation effects at the cost of a proper 
treatment of electron spin.  Spin symmetry is restored in CI methods, but these can only 
be applied to finite clusters owing to the large number of electronic configurations 
encountered in that approach.  CI calculations on the pair of \Mnoct octahedra in the 
polaron in \LaCa show that the overall spin configuration is a linear 
combination of several determinants \cite{Patterson03} with a magnetic moment on the central oxygen ion.
The \LaCa polaron electronic 
structure therefore resembles that in $\alpha$'-\NaVO \cite{Suaud00}; it is unlike the electronic structure 
which results in FM coupling in \Lap, which is very well 
described by a single spin configuration \cite{Nicastro02}.

An obvious question which arises within the polaron picture is, 'What happens when 
electrons are added to the x = 1/2 phase?'  That is, 'What is the electronic structure 
predicted to be in the FM region of the phase diagram with 0.2 $<$ x $<$ 0.5?'  Since the 
bottom of the conduction band in \LaCa is comprised largely of vacant oxygen
2p states on the \ooneion ion in the polaron, one would naturally expect to add the extra 
electron here.  The spin density plots 
in Figs. \ref{fig:fig3} and \ref{fig:fig4} show that OO in the x = 1/2 compounds is \textit{parallel} to the polaron axis, 
whereas it is T-shaped in \La (long Mn-O bonds in 
JT distorted octahedra in \La are \textit{perpendicular} to each other).
If the extra electron is added to the polaronic oxygen
site, it is expected that there will be reorientation of \eg orbitals (Fig. \ref{fig:fig1}c) and adjustment of Mn-O-Mn 
bond lengths.  OO in the vicinity of the added electron is expected to resemble 
that in \La and one might expect 
to nucleate a small FM patch as further electrons are added.  

\acknowledgments
The authors wish to acknowledge discussions with W. Mackrodt, P.G. Radaelli, M. 
Towler, V. Ferrari and P. Midgley.  This work was supported by Enterprise Ireland 
under grant number SC/00/267.



\begin{thebibliography}{42}
\bibitem{structures}
The P21/m structure used for \LaCa was obtained by x-ray diffraction at 20 K (Table I 
Ref. \cite{Radaelli97}). The Pnma Wigner crystal structure used for \LaCathree was obtained by 
powder neutron diffraction at 1.5 K
(Table V of Ref. \cite{Radaelli99}).
\end{thebibliography}
\end{document}